\documentclass{ws-ijmpa}

\usepackage[super]{cite}
\usepackage{xcolor}
\usepackage[hypertexnames=false]{hyperref}
\hypersetup{colorlinks=false,allbordercolors=blue,pdfborderstyle={/S/U/W 1}}
\usepackage{graphicx}
\usepackage{amsmath,amssymb}
\usepackage{booktabs}
\usepackage{subcaption}

\begin{document}

\markboth{X.K.~Li et al.}{Quantum entanglement in tau pairs}

\catchline{}{}{}{}{}

\title{Probing quantum entanglement in $\tau^+\tau^-$ pairs via the $\pi\pi$ channel at STCF}

\author{Xiaokang Li}
\address{Peking University, Beijing, P. R. China\\
lixk@stu.pku.edu.cn}

\author{Chentao Bao}
\address{Zhejiang University, Hangzhou, P. R. China}

\author{Hai Chen}
\address{Zhejiang University, Hangzhou, P. R. China}

\author{Mingyi Liu}
\address{University of Science and Technology of China, Hefei, P. R. China}

\author{Dayong Wang}
\address{Peking University, Beijing, P. R. China}

\maketitle

\begin{history}
\received{}
\end{history}

\begin{abstract}
Quantum entanglement and Bell-inequality violation in $\tau^+\tau^-$ pairs provide a sensitive probe of quantum correlations in high-energy interactions. We present a feasibility study of $e^+e^-\to\tau^+\tau^-$ at the proposed Super Tau-Charm Facility (STCF) based on full Monte Carlo simulation at $\sqrt{s}=7$\,GeV, focusing on the $\pi\pi$ channel ($\tau^\pm\to\pi^\pm\nu$), which offers the maximal spin-analyzing power $|\kappa|=1$ and the simplest final-state topology for validating the quantum-tomography framework. We establish a consistency chain from the tree-level QED prediction through truth-level and detector-level reconstruction, yielding a reconstructed concurrence of $\mathcal{C}=0.279\pm0.007$ with the good-solution approach. A complementary full-simulation study of the $\rho\rho$ channel is also briefly reported. These results demonstrate that the STCF can provide a competitive platform for precision studies of quantum correlations in $\tau$-lepton pairs.
\end{abstract}

\keywords{Quantum entanglement; Bell inequality; $\tau$ lepton; spin correlation; Super Tau-Charm Facility.}

\section{Introduction}

Quantum mechanics and its relativistic extension into quantum field theory provide the most successful framework for describing nature at the smallest scales. Yet the features that most sharply distinguish quantum mechanics from classical theories have played a limited role in collider phenomenology. Quantum information science, by contrast, has developed these into laboratory resources on atomic, photonic and solid-state platforms. Central to this effort is quantum entanglement, in which the states of two subsystems that have interacted in the past cannot in general be described independently, and the correlations between them exceed anything reproducible by a classical mixture of product states. A stronger statement is Bell-inequality violation (BIV), which rules out any local hidden-variables description.\cite{bell_chsh} 

High-energy collisions also provide a natural arena for such tests.\cite{barr_review} Pair-produced particles emerge in quantum-correlated spin configurations fixed by the production dynamics and angular-momentum conservation, allowing quantum correlations to be probed at energy scales of $\mathcal{O}(\text{GeV})$ and above, and under weak and strong interactions distinct from the electromagnetic settings of traditional tests. The key experimental technique is quantum state tomography, in which the spin density matrix of unstable particles is reconstructed from the angular distributions of their decay products.\cite{tomography_bernal} Following this approach, spin entanglement in top-quark pair production has been observed by the ATLAS\cite{atlas_top} and CMS\cite{cms_top,cms_top_lj} Collaborations at the LHC. Phenomenological studies have extended the program to $\tau$-lepton pairs at $e^+e^-$ colliders\cite{ehat_belle2,fabbrichesi_fccee,han_bepc_tautau,ma_cepc_bell,altakach_polarized,yang_stcf_tautau} and at the LHC using machine-learning-based neutrino reconstruction,\cite{zhang_lhc_tautau} and have demonstrated the potential of entanglement observables to constrain physics beyond the Standard Model.\cite{fabbrichesi_bsm}

Among the systems available at colliders, $\tau^+\tau^-$ pairs occupy a special position. The $\tau$ is the only lepton heavy enough to decay hadronically and the parity-violating nature of its weak decays makes the decay products particularly clean spin analyzers, with their angular distributions directly encoding the parent $\tau$ polarization. In particular, the two-body decay $\tau\to\pi\nu$ has the maximal spin analyzing power $|\kappa|=1$, meaning the pion direction provides unambiguous access to the $\tau$ spin state and makes the $\pi\pi$ channel an ideal benchmark for quantum tomography.\cite{ehat_belle2,jadach_tauola}

The Super Tau-Charm Facility (STCF)\cite{stcf_cdr} is a next-generation $e^+e^-$ collider designed to operate at center-of-mass energies from 2 to 7\,GeV with a peak luminosity of $0.5\times10^{35}~\text{cm}^{-2}\text{s}^{-1}$, expected to produce approximately $1.9\times10^{9}$ $\tau$ pairs per year at $\sqrt{s}=7$\,GeV. In this proceeding, we report a feasibility study of probing quantum entanglement and BIV in $e^+e^-\to\tau^+\tau^-$ at the STCF via the $\pi\pi$ channel ($\tau^\pm\to\pi^\pm\nu$), with results from an ongoing full-simulation study of the $\rho\rho$ channel briefly mentioned.

\section{Quantum Entanglement of $\tau$-Pair System}
\label{sec:formalism}

\subsection{Theoretical framework}

The spin state of a $\tau^+\tau^-$ pair produced in $e^+e^-$ collisions can be described by a $4\times4$ spin density matrix (SDM), which can be decomposed as
\begin{equation}
\rho_{\tau^+\tau^-} = \frac{1}{4}\biggl[
  \mathbf{1}\otimes\mathbf{1}
  + \sum_i B_i^+\,(\sigma_i\otimes\mathbf{1})
  + \sum_j B_j^-\,(\mathbf{1}\otimes\sigma_j)
  + \sum_{i,j} C_{ij}\,(\sigma_i\otimes\sigma_j)
\biggr],
\label{eq:sdm}
\end{equation}
where $\sigma_i$ are the Pauli matrices, $B_i^\pm$ denote the components of the polarization vectors of $\tau^\pm$, and $C_{ij}$ denote the components of the spin-correlation matrix $\mathbf{C}$. The indices $i,j$ run over the right-handed helicity basis $\{\hat{n},\hat{r},\hat{k}\}$ defined in the $\tau^+\tau^-$ center-of-mass frame, with $\hat{k}$ along the $\tau^+$ direction and
\begin{equation}
\hat{n} = \frac{\hat{p}\times\hat{k}}{\sin\theta}\,,
\qquad
\hat{r} = \frac{\hat{p}-\hat{k}\cos\theta}{\sin\theta}\,,
\label{eq:nrk}
\end{equation}
where $\hat{p}$ is the beam direction and $\theta$ is the $\tau$ scattering angle.

Two observables are employed to test quantum correlations in the $\tau^+\tau^-$ system:
\begin{itemlist}
\item \textit{Concurrence.} A bipartite state is entangled if its density matrix cannot be written as a convex combination of product states. The concurrence\cite{wootters_concurrence} provides a faithful measure of entanglement for a two-qubit system:
\begin{equation}
\mathcal{C}[\rho] = \max\{0,\,\lambda_1-\lambda_2-\lambda_3-\lambda_4\} \in [0,1]\,,
\label{eq:concurrence}
\end{equation}
where $\lambda_1\ge\lambda_2\ge\lambda_3\ge\lambda_4$ are the eigenvalues, in decreasing order, of the matrix
\begin{equation}
\mathbf{R} = \sqrt{\sqrt{\rho}\,\tilde{\rho}\,\sqrt{\rho}}\,, \quad \text{with} \quad \tilde{\rho} = (\sigma_2\otimes\sigma_2)\,\rho^*\,(\sigma_2\otimes\sigma_2)\,.
\label{eq:R_matrix}
\end{equation}
A state is separable if and only if $\mathcal{C}=0$, and maximally entangled when $\mathcal{C}=1$.

\item \textit{Bell-inequality violation.} The Clauser-Horne-Shimony-Holt (CHSH) inequality\cite{bell_chsh} provides a sharper test of quantum correlations, as its violation rules out any description in terms of local hidden variables. For a two-qubit system, the maximal CHSH violation is captured by the Horodecki condition\cite{horodecki_biv}: defining $\mathbf{M} = \mathbf{C}^T\mathbf{C}$ with eigenvalues $m_1\ge m_2\ge m_3$ in decreasing order, the state violates the CHSH inequality if and only if
\begin{equation}
\mathfrak{m}_{12}[\mathbf{C}] \equiv m_1 + m_2 > 1\,.
\label{eq:m12}
\end{equation}
In the collider context, $\mathfrak{m}_{12}>1$ is not a strict Bell test, as it relies on the assumption that the optimal measurement directions can be chosen event by event; it nevertheless serves as a powerful witness for nonlocal correlations accessible through quantum tomography.
\end{itemlist}

\begin{figure}[!ht]
\centerline{
  \includegraphics[width=0.32\textwidth]{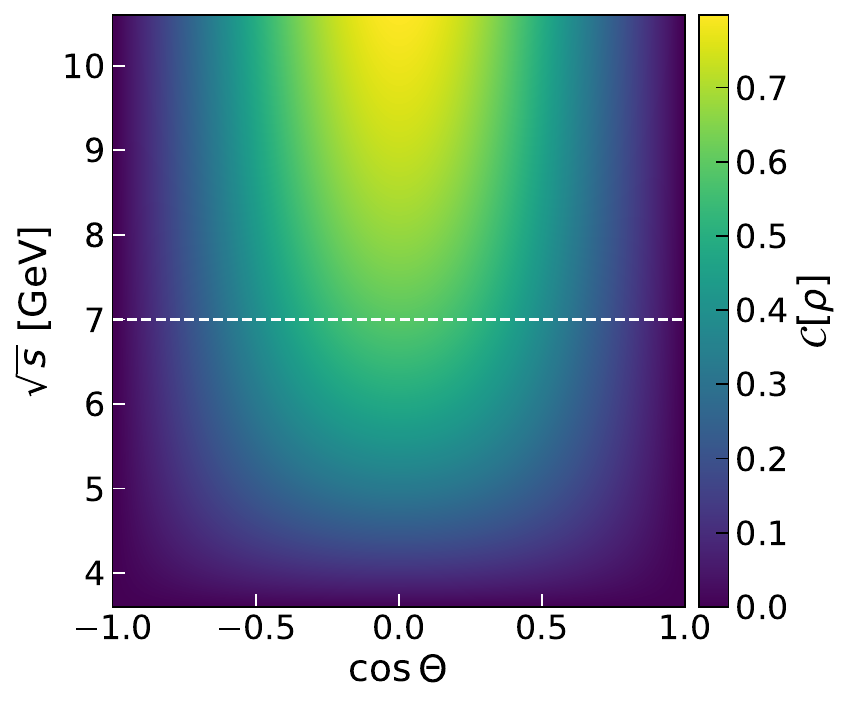}
  \includegraphics[width=0.32\textwidth]{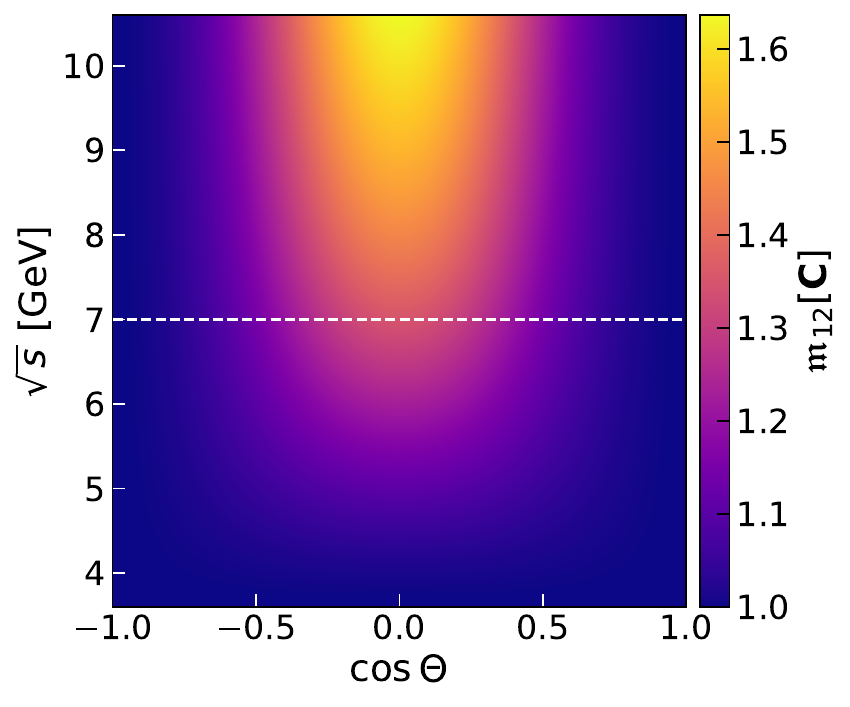}
  \includegraphics[width=0.32\textwidth]{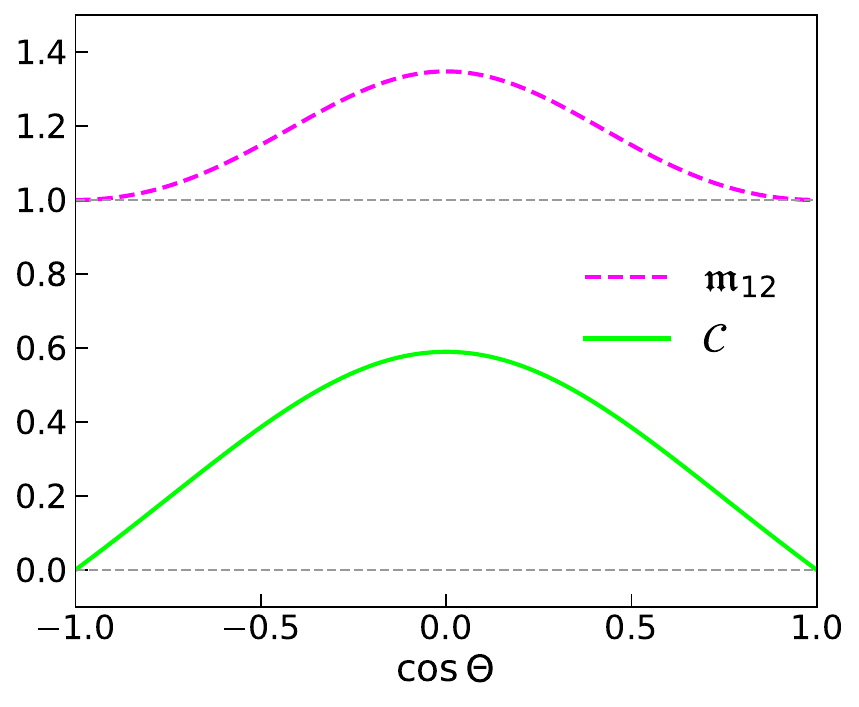}
}
\caption{SM predictions for $\mathcal{C}$ (left), $\mathfrak{m}_{12}$ (middle), and $\mathcal{C}$ and $\mathfrak{m}_{12}$ at $\sqrt{s}=7$\,GeV (right).}
\label{fig:sm_prediction}
\end{figure}

For the tree-level QED process $e^+e^-\to\gamma^*\to\tau^+\tau^-$, the SM predicts $B_i^\pm=0$ and a diagonal correlation matrix $C_{ij}$, yielding the analytical expressions\cite{ehat_belle2}
\begin{equation}
\mathcal{C} =
  \frac{(s-4m_\tau^2)\sin^2\theta}
  {4m_\tau^2\sin^2\theta + s(\cos^2\theta+1)}\,,    
\end{equation}
\begin{equation}
\mathfrak{m}_{12} = 1 +
  \biggl[\frac{(s-4m_\tau^2)\sin^2\theta}
  {4m_\tau^2\sin^2\theta + s(\cos^2\theta+1)}
  \biggr]^2,
\label{eq:sm_prediction}
\end{equation}
as functions of the center-of-mass energy squared $s$ and the $\tau$ scattering angle $\theta$, as shown in Fig.~\ref{fig:sm_prediction}. Both entanglement and BIV become more pronounced at higher energies and smaller $|\cos\theta|$. We choose $\sqrt{s}=7$\,GeV for the entanglement study, as it offers the largest center-of-mass energy accessible at the STCF and thus maximizes the expected entanglement signal.

\subsection{Experimental reconstruction}

Experimentally, the SDM elements are extracted through the polarimeter-vector method\cite{ehat_belle2}. For each $\tau$ decay, a polarimeter vector $\vec{h}^\pm$ is constructed in the respective $\tau^\pm$ rest frame from the kinematics of the decay products. The polarization and correlation coefficients are obtained as
\begin{align}
B_i^\pm &= \frac{3}{\kappa_\pm}\cdot\frac{1}{\sigma}\int\mathrm{d}\Omega^\pm\,\frac{\mathrm{d}\sigma}{\mathrm{d}\Omega^\pm}(\vec{h}^\pm\cdot\hat{e}_i) = \frac{3}{\kappa_\pm}\,\langle h_i^\pm\rangle\,, \label{eq:B_extract} \\
C_{ij} &= \frac{9}{\kappa_-\kappa_+}\cdot\frac{1}{\sigma}\iint\mathrm{d}\Omega^-\mathrm{d}\Omega^+\,\frac{\mathrm{d}^2\sigma}{\mathrm{d}\Omega^-\mathrm{d}\Omega^+}(\vec{h}^-\cdot\hat{e}_i)(\vec{h}^+\cdot\hat{e}_j) = \frac{9}{\kappa_-\kappa_+}\,\langle h_i^-\,h_j^+\rangle\,, \label{eq:C_extract}
\end{align}
where $h_i^\pm = \vec{h}^\pm\cdot\hat{e}_i$ ($\hat{e}_i = \hat{n},\hat{r},\hat{k}$) and $\kappa_\pm$ is the spin analyzing power of the decay channel.

The polarimeter vector depends on the $\tau$ decay mode\cite{jadach_tauola}. For the two-body decay $\tau^\pm\to\pi^\pm\nu$, it takes the simplest form,
\begin{equation}
\vec{h}^\pm = -\hat{n}_{\pi^\pm}\,, \qquad \kappa_\pm = \pm 1\,,
\label{eq:pol_pipi}
\end{equation}
where $\hat{n}_{\pi^\pm}$ is the unit vector along the pion direction in the $\tau^\pm$ rest frame. The analyzing power $|\kappa|=1$ is maximal, meaning the pion direction fully encodes the $\tau$ spin information.

\section{The STCF Detector and Monte Carlo Simulation}
\label{sec:detector}

The STCF\cite{stcf_cdr} is a proposed next-generation $e^+e^-$ collider designed to operate at $\sqrt{s}=2$ to 7\,GeV, with a peak luminosity of $0.5\times10^{35}\,\text{cm}^{-2}\text{s}^{-1}$ at $\sqrt{s}=4.0$\,GeV. From inside out, the baseline detector consists of an inner tracker system using $\mu$-RWELL-based gaseous detectors or MAPS-based silicon pixel detectors, a main drift chamber, a particle identification system in both barrel and endcap regions, an electromagnetic calorimeter (EMC), a 1~T superconducting solenoid, and a muon detector. This configuration provides the broad solid-angle coverage, high tracking and photon efficiency and robust hadron identification required for the $\tau$-pair analyses presented here. At $\sqrt{s}=7$\,GeV, about $1.9\times10^{9}$ $\tau$ pairs are expected per year, providing a sample well matched to precision entanglement studies.

Monte Carlo samples are produced within the STCF offline software framework OSCAR\cite{oscar}, with a full Geant4-based simulation\cite{geant4}, digitization and reconstruction chain. Signal $e^+e^-\to\tau^+\tau^-$ events are generated at leading order with MadGraph5\_aMC@NLO\cite{madgraph}, with $\tau$ spin correlations and the dominant decay modes handled through its Decay Chain module and the \texttt{taudecay\_UFO} library; subdominant $\tau$ decays and initial-state radiation are simulated by Pythia~8.306\cite{pythia8}. A sample of 30 million inclusive $\tau$-pair events is used for the $\pi\pi$ channel analysis.

\section{Analysis of the $\pi\pi$ Channel}
\label{sec:pipi_analysis}

The $\pi\pi$ channel refers to $e^+e^-\to\tau^+\tau^-$ with both $\tau$ leptons decaying via $\tau^\pm\to\pi^\pm\nu$. Although its combined branching fraction is only about $1.4\%$, the maximal spin analyzing power and the minimal final-state topology make it an ideal benchmark for validating the quantum-tomography methodology on signal Monte Carlo. The analysis presented here establishes the consistency of the full reconstruction chain against the tree-level QED prediction; a full background and random-solution treatment is left to future work.

\subsection{$\tau$ momentum reconstruction}
\label{sec:tau_reco}

The reconstruction of the $\tau$ momentum is a prerequisite for building the polarimeter vectors and extracting the SDM. In the $\tau^+\tau^-$ center-of-mass frame, and neglecting initial-state-radiation (ISR), the invariant-mass conditions $(p_\pi + p_\nu)^2 = m_\tau^2$ and $p_\nu^2 = 0$, together with the initial-state four-momentum constraint $\sum_i p_i = (E_\text{cms},\vec{0})$, fix the $\tau^+$ energy to $E_\tau = E_\text{cms}/2$ and hence the momentum magnitude $|\vec{k}| = \sqrt{E_\tau^2 - m_\tau^2}$. Its direction $\hat{k}$ is constrained by\cite{Belle:2021ybo}
\begin{equation}
\cos\theta_i = \frac{2 E_i E_\tau - m_i^2 - m_\tau^2}{2|\vec{k}|\,|\vec{p}_i|}\,, \qquad i = \pi^-, \pi^+\,,
\label{eq:cos_theta_k}
\end{equation}
which specifies the angle between $\hat{k}$ and each observed pion momentum but leaves a two-fold ambiguity: the two solutions are symmetric with respect to the plane containing the two pion momenta.

In principle, the ambiguity can be resolved using the $\tau$ decay vertex, since the correct solution must lie along the true $\tau$ flight direction. At the STCF, however, the bunch length ($\sigma_z\sim10$\,mm) is much larger than the $\tau$ decay length ($L_\tau\sim140\,\mu$m at $\sqrt{s}=7$\,GeV), and the current tracking resolution ($\sigma_{D_0}\sim 20\,\mu$m and $\sigma_{Z_0}\sim100\,\mu$m, where $D_0$ and $Z_0$ denote the transverse and longitudinal impact parameters, respectively) is insufficient to reconstruct the $\tau$ decay vertex with the required precision\cite{stcf_cdr}. The ambiguity therefore cannot be eliminated at the present stage. For the validation presented below, the \textit{good solution} --- the one closer to the truth $\tau$ direction --- is retained in order to isolate acceptance and reconstruction effects from the ambiguity itself, linking the tree-level QED prediction, the full-phase-space (full-PHSP) truth result, the truth result after selection and the good-solution reconstruction. A truth-independent resolution of the ambiguity, e.g.\ by random choice between the two solutions, is deferred to future work.

\subsection{Event selection and tomography results}
\label{sec:pipi_results}

Events are selected by requiring exactly two oppositely charged tracks ($D_0<1$\,cm, $Z_0<10$\,cm, $|\cos\theta|<0.95$) identified as pions. The subsequent validation proceeds in two steps using the signal sample alone.

\textit{Full phase space (truth level).} As a first cross-check, the algorithm is run on truth information without any detector effects or acceptance cuts. The polarimeter vectors are built directly from the truth-level pion directions in the corresponding $\tau^\pm$ rest frames, and $B_i^\pm$, $C_{ij}$ are extracted via Eqs.~(\ref{eq:B_extract}) and~(\ref{eq:C_extract}). The reconstructed $\mathcal{C}=0.314\pm0.004$ and $\mathfrak{m}_{12}=0.885\pm0.009$ agree closely with the tree-level QED prediction at $\sqrt{s}=7$\,GeV, $\mathcal{C}=0.329$ and $\mathfrak{m}_{12}=0.906$, with the small residual differences consistent with ISR effects included in the simulation.

\begin{figure}[!ht]
\centerline{\includegraphics[width=\textwidth]{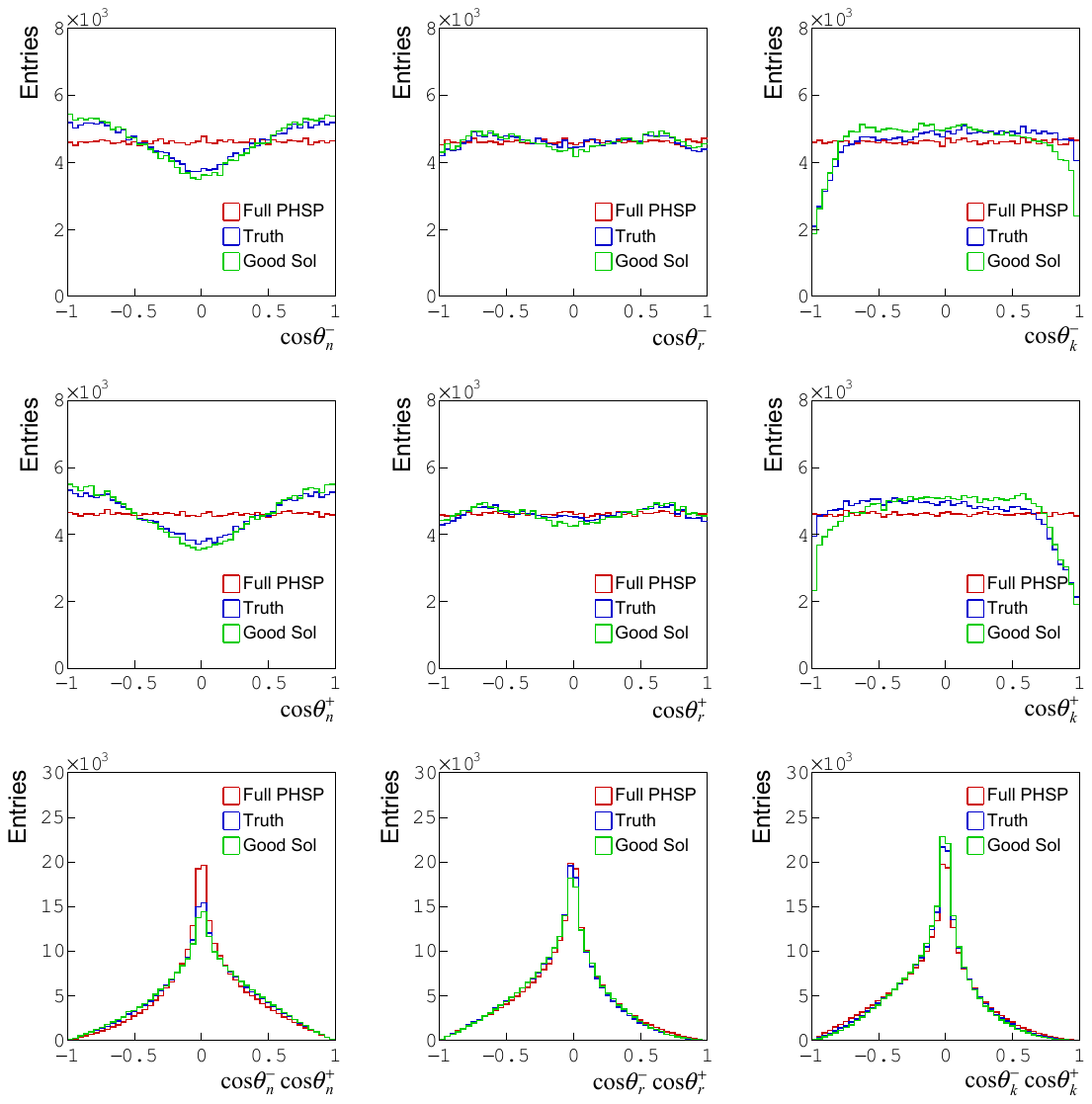}}
\caption{(Color online) Distributions of $\cos\theta_i^\pm=-h_i^\pm$ (top two rows) and $\cos\theta_i^-\cos\theta_j^+=h_i^- h_j^+$ (bottom one row) along the diagonal $\{\hat{n},\hat{r},\hat{k}\}$ directions in the $\pi\pi$ channel, compared among the full-PHSP truth sample (red histogram), the in-acceptance truth sample (blue histogram) and the corresponding good-solution reconstruction (green histogram).}
\label{fig:BC_pipi}
\end{figure}

\textit{Truth versus good solution.} The detector response and the two-fold ambiguity are then introduced. Each event passing the selection is analyzed twice: once using the $\tau$ truth direction (truth), and once using the reconstructed solution closer to it (good solution). As shown in Fig.~\ref{fig:BC_pipi}, the two largely agree across the polarimeter angular distributions, while both shift relative to the full-PHSP distributions owing to acceptance losses from the $|\cos\theta|<0.95$ selection. The extracted concurrence values, $\mathcal{C}^{\text{Truth}}=0.296\pm0.007$ and $\mathcal{C}^{\text{Good}}=0.279\pm0.007$, are consistent within uncertainties. The resulting consistency chain --- tree-level QED $\to$ full-PHSP truth $\to$ truth-in-acceptance $\to$ good solution --- confirms that the tomography algorithm faithfully recovers the entanglement signal when the correct $\tau$ direction is known.

\section{Summary}
\label{sec:summary}

We have presented a feasibility study of quantum entanglement and Bell-inequality violation in $e^+e^-\to\tau^+\tau^-$ at the STCF via the $\pi\pi$ channel, using full Monte Carlo simulation at $\sqrt{s}=7$\,GeV. The tomography algorithm is validated through a consistency chain from the tree-level QED prediction to the good-solution reconstruction, yielding $\mathcal{C}=0.279\pm0.007$ and confirming that the entanglement signal is faithfully recovered. A complementary full-simulation study of the $\rho\rho$ channel yields $\mathcal{C}=0.34\pm0.02$ for $|\cos\theta|<0.4$ and $\mathfrak{m}_{12}=1.19\pm0.07$ for $|\cos\theta|<0.1$.

The main limitation is the two-fold ambiguity in $\tau$ momentum reconstruction, which cannot be resolved with the current vertex resolution. Ongoing work includes a random-solution treatment and full background evaluation for the $\pi\pi$ channel and extending the analysis to the $\tau\to 3\pi\nu$ mode, whose three-prong topology may help lift this ambiguity. These results confirm that the STCF can serve as a competitive platform for precision quantum-correlation studies in $\tau$-lepton pairs.

\section*{Acknowledgments}
We thank the supercomputing center of Lanzhou University for their strong support. We are grateful to Yu Zhang, convener of the STCF basic symmetries physics simulation group, for valuable discussions and support. We also thank the STCF hardware and software working groups for their contributions to the detector simulation and offline software framework used in this work. This work is supported by the National Key R\&D Program of China under Contracts No.~2022YFA1602200 and No.~2023YFA1607200; the National Natural Science Foundation of China (NSFC) under Contracts Nos.~12341501, 12341503, 12341504; and the International Partnership Program of the Chinese Academy of Sciences under Grant No.~211134KYSB20200057. We thank the Hefei Comprehensive National Science Center for their strong support on the STCF key technology research project.

\section*{ORCID}
Xiaokang Li -- https://orcid.org/0009-0008-8476-3932\\
Chentao Bao -- https://orcid.org/0009-0009-3962-2775\\
Hai Chen -- https://orcid.org/0000-0003-4195-9966\\
Mingyi Liu -- https://orcid.org/0000-0002-0236-5404\\
Dayong Wang -- https://orcid.org/0000-0002-9013-1199


\begin{thebibliography}{99}

\bibitem{bell_chsh}
J.~F.~Clauser, M.~A.~Horne, A.~Shimony and R.~A.~Holt,
{\it Phys. Rev. Lett.} {\bf 23}, 880 (1969),
\url{https://doi.org/10.1103/PhysRevLett.23.880}.

\bibitem{barr_review}
A.~J.~Barr, M.~Fabbrichesi, R.~Floreanini, E.~Gabrielli and L.~Marzola,
{\it Prog. Part. Nucl. Phys.} {\bf 139}, 104134 (2024),
\url{https://doi.org/10.1016/j.ppnp.2024.104134}.

\bibitem{tomography_bernal}
A.~Bernal,
{\it Phys. Rev. D} {\bf 109}, 116007 (2024),
\url{https://doi.org/10.1103/PhysRevD.109.116007}.

\bibitem{atlas_top}
ATLAS Collab. (G.~Aad {\it et al.}),
{\it Nature} {\bf 633}, 542 (2024),
\url{https://doi.org/10.1038/s41586-024-07824-z}.

\bibitem{cms_top}
CMS Collab. (A.~Hayrapetyan {\it et al.}),
{\it Rep. Prog. Phys.} {\bf 87}, 117801 (2024),
\url{https://doi.org/10.1088/1361-6633/ad7e4d}.

\bibitem{cms_top_lj}
CMS Collab. (A.~Hayrapetyan {\it et al.}),
{\it Phys. Rev. D} {\bf 110}, 112016 (2024),
\url{https://doi.org/10.1103/PhysRevD.110.112016}.

\bibitem{ehat_belle2}
K.~Ehat\"aht, M.~Fabbrichesi, L.~Marzola and C.~Veelken,
{\it Phys. Rev. D} {\bf 109}, 032005 (2024),
\url{https://doi.org/10.1103/PhysRevD.109.032005}.

\bibitem{fabbrichesi_fccee}
M.~Fabbrichesi and L.~Marzola,
{\it Phys. Rev. D} {\bf 110}, 076004 (2024),
\url{https://doi.org/10.1103/PhysRevD.110.076004}.

\bibitem{han_bepc_tautau}
T.~Han, M.~Low and Y.~Su,
{\it JHEP} {\bf 10}, 217 (2025),
\url{https://doi.org/10.1007/JHEP10(2025)217}.

\bibitem{ma_cepc_bell}
K.~Ma and T.~Li,
{\it Chin. Phys. C} {\bf 48}, 103105 (2024),
\url{https://doi.org/10.1088/1674-1137/ad58f5}.

\bibitem{altakach_polarized}
M.~M.~Altakach, P.~Lamba, F.~Maltoni and K.~Sakurai,
arXiv:2601.09558 [hep-ph],
\url{https://arxiv.org/abs/2601.09558}.

\bibitem{yang_stcf_tautau}
B.~Yang, Y.~Zhang, Z.~S.~Wang and X.~Zhou,
arXiv:2603.05846 [hep-ph],
\url{https://arxiv.org/abs/2603.05846}.

\bibitem{zhang_lhc_tautau}
Y.~Zhang {\it et al.},
arXiv:2504.01496 [hep-ph],
\url{https://arxiv.org/abs/2504.01496}.

\bibitem{fabbrichesi_bsm}
M.~Fabbrichesi, R.~Floreanini, E.~Gabrielli and L.~Marzola,
{\it Eur. Phys. J. C} {\bf 83}, 162 (2023),
\url{https://doi.org/10.1140/epjc/s10052-023-11307-2}.

\bibitem{jadach_tauola}
S.~Jadach, J.~H.~K\"uhn and Z.~Was,
{\it Comput. Phys. Commun.} {\bf 64}, 275 (1990),
\url{https://doi.org/10.1016/0010-4655(91)90038-M}.

\bibitem{stcf_cdr}
M.~Achasov {\it et al.},
{\it Front. Phys. (Beijing)} {\bf 19}, 14701 (2024),
\url{https://doi.org/10.1007/s11467-023-1333-z}.

\bibitem{wootters_concurrence}
W.~K.~Wootters,
{\it Phys. Rev. Lett.} {\bf 80}, 2245 (1998),
\url{https://doi.org/10.1103/PhysRevLett.80.2245}.

\bibitem{horodecki_biv}
R.~Horodecki, P.~Horodecki and M.~Horodecki,
{\it Phys. Lett. A} {\bf 200}, 340 (1995),
\url{https://doi.org/10.1016/0375-9601(95)00214-N}.

\bibitem{oscar}
W.~H.~Huang {\it et al.},
{\it J. Instrum.} {\bf 18}, P03004 (2023),
\url{https://doi.org/10.1088/1748-0221/18/03/P03004}.

\bibitem{geant4}
S.~Agostinelli {\it et al.} [GEANT4 Collaboration],
{\it Nucl. Instrum. Meth. A} {\bf 506}, 250 (2003),
\url{https://doi.org/10.1016/S0168-9002(03)01368-8}.

\bibitem{madgraph}
J.~Alwall {\it et al.},
{\it JHEP} {\bf 07}, 079 (2014),
\url{https://doi.org/10.1007/JHEP07(2014)079}.

\bibitem{pythia8}
C.~Bierlich {\it et al.},
{\it SciPost Phys. Codeb.}, 8 (2022),
\url{https://doi.org/10.21468/SciPostPhysCodeb.8}.

\bibitem{Belle:2021ybo}
Belle Collab. (K.~Inami {\it et al.}),
{\it JHEP} {\bf 04}, 110 (2022),
\url{https://doi.org/10.1007/JHEP04(2022)110}.

\end{thebibliography}
\end{document}